\documentclass[11pt,reqno]{amsart}
\usepackage{amsmath}
\usepackage{amsfonts}
\usepackage{amsthm}

\theoremstyle{definition}
\newtheorem{definition}{Definition}[section]

\newtheorem{remark}[definition]{Remark}
\newtheorem{remarks}[definition]{Remarks}

\newtheorem{lemma}[definition]{Lemma}

\newtheorem{theorem}[definition]{Theorem}

\newtheorem{consequence}[definition]{Consequence}
\newtheorem{conjecture}[definition]{Conjecture}

\theoremstyle{remark}

\newcommand\F{{\mathcal F}}

\newcommand\N{{\mathbb N}}
\newcommand\C{{\mathbb C}}
\newcommand\dz{{\frac{d^2z}{(1+\abs{z}^2)^2}}}
\renewcommand\SS{{\mathbb S}^2}

\newcommand\ie{{i.$\,$e.\ }}

\def\downto{{\mathchoice
{\raise.25ex\hbox{$\,\scriptstyle\searrow\;$}} 
{\raise.25ex\hbox{$\,\scriptstyle\searrow\;$}} 
{\raise.25ex\hbox{$\scriptscriptstyle\searrow$}} 
{\raise.25ex\hbox{$\scriptscriptstyle\searrow$}} 
}}
\def\upto{{\mathchoice
{\raise.25ex\hbox{$\,\scriptstyle\nearrow\;$}} 
{\raise.25ex\hbox{$\,\scriptstyle\nearrow\;$}} 
{\raise.25ex\hbox{$\scriptscriptstyle\nearrow$}} 
{\raise.25ex\hbox{$\scriptscriptstyle\nearrow$}} 
}}

\newcommand\abs[1]{{\left\vert{#1}\right\vert}}
\newcommand\norm[1]{{{\vert\mkern-1mu\vert}{#1}{\vert\mkern-1mu\vert}}}
\newcommand\nnorm[1]{{{\vert\mkern-2mu\vert\mkern-2mu\vert}{#1}{\vert\mkern-2mu\vert\mkern-2mu\vert}}}
\newcommand\eq[1]{(\ref{eq:#1})}
\newcommand\ol{\overline}
\newcommand\conf[1]{{(1+\abs{z}^2)^{#1}}}
\newcommand\reg[1]{{(\abs{\phi(z)}^2+\epsilon)^{#1}}}

\renewcommand\MR[1]{}

\title[A lower bound for the Wehrl entropy of quantum spin]{A lower 
       bound for the Wehrl entropy of quantum spin with sharp high-spin asymptotics}
\author{Bernhard G. Bodmann}

\begin{document}

\maketitle
\vspace*{-0.6cm}
\centerline{{\small Department of Physics, Princeton University, Princeton, NJ 08544}}
\centerline{{\small bgb@math.princeton.edu}}

\begin{abstract}
A lower bound for the Wehrl entropy of
a single quantum spin is derived. The 
high-spin asymptotics of this bound coincides with
Lieb's conjecture up to, but not including, terms of first and 
higher order in the inverse spin quantum number. The result 
presented here may be seen as complementary to the verification of
the conjecture in cases of lowest spin by Schupp 
[Commun.\ Math.\ Phys.\ {\bf 207} (1999), 481].
The present result for the Wehrl-entropy 
is obtained from interpolating a sharp norm bound 
that also implies a sharp lower bound for the so-called 
R\'enyi-Wehrl entropy with certain indices that 
are evenly spaced by half of the inverse spin quantum 
number.
\end{abstract}

\section{Introduction}

Among the many facets of entropy, Wehrl \cite{Weh91} investigated the relation 
between the usual Gibbs-Boltzmann-Shannon entropy of a density in phase space 
and its quantum analogue, commonly known as the von~Neumann entropy of a quantum 
state. The question how to recover one from the other in the sense of
a classical limit led Wehrl to construct a quasi-classical entropy of quantum states 
\cite{Weh79}. This hybrid construction derived from so-called Glauber coherent states 
is non-negative, or more accurately, bounded below by the von~Neumann entropy. Described 
in physical terms, the presence of an uncertainty principle in the quasi-classical 
formulation imposes the lower bound on the usual classical entropy.

Wehrl conjectured that the minimum value of his quasi-classical entropy 
is assumed whenever the quantum system is in a coherent state. 
This was proved by Lieb \cite{Lie78} who established sharp norm bounds for the 
Bargmann transform \cite{Bar61}, an isometry between Schr\"odinger's
position-space representation and a phase-space representation associated with 
Glauber coherent states. 
In addition, Lieb \cite{Lie78} suggested that coherent states should also be the 
minimizers for the Wehrl entropy of quantum spin systems. The
quasi-classical entropy for a single quantum spin
is derived from coherent states that contain a highest-weight 
vector in a finite-dimensional Hilbert space
carrying a unitary irreducible representation of SU(2). 
In this finite-dimensional setting, one could have hoped for a simpler way to prove the corresponding
bound on the Wehrl entropy compared to the effort and ingenuity in the proof given by Lieb for the case 
of Glauber coherent states. In fact, this would have provided an alternative to Lieb's original proof via the 
usual group-contraction procedure, in which Glauber coherent states are obtained as
the high-spin limit of SU(2) coherent states \cite{ACGT72}. However, despite a lot of attempts,
such a proof did not materialize.

Perhaps the simplest proof of Wehrl's original conjecture was given by Carlen \cite{Car91}
and in a similar argument by Luo \cite{Luo00}.
It reduces to an equality between the norm of functions in the phase-space
representation and the Dirichlet form of the Laplacian, together with Gross' logarithmic
Sobolev inequality \cite{Gro75}. Unfortunately, the combination of such identities on the
sphere does not yield the desired bound for the Wehrl entropy of a quantum spin. The reason 
is that the logarithmic Sobolev inequality on the sphere \cite{MW82} has the constant as its
optimizer, a function that is not contained in the phase-space representation
for spin.
 
A few years ago, Lieb's conjecture was verified for the non-trivial case of
spin quantum numbers $j=1$ and $j=3/2$ \cite{Sch99}. This was done by explicit 
calculation of the entropy in a geometric parametrization of quantum states that 
also appeared in \cite{Lee88}, combined with elementary inequalities involving 
chordal distances of points on the sphere. Here, we derive a lower bound on the Wehrl 
entropy that coincides with Lieb's conjecture up to first order in the inverse of the 
spin quantum number. To this end, we modify Carlen's approach and its implicit idea that
\textit{the Wehrl entropy bound should be a result of the hypercontractivity properties
of the Berezin transform}. In physical terms, increasing the value of
Planck's constant adds fluctuations and smoothes the phase-space density associated with a 
given quantum state. As usual, we evaluate the smoothness of a density by the 
$s$-dependence of its $s$-norm, and understand the Wehrl entropy as related to
the infinitesimal change at $s=1$. Carlen's technique amounts to establishing bounds
between norms for continuously tuned values of Planck's constant. Its analogue in the 
setting of spin is the inverse of the spin quantum number that cannot be tuned continuously, 
which is ultimately the reason why the entropy bound obtained here is not sharp. Fortunately, 
for large spin quantum numbers this conceptual problem is less and less relevant which 
leads to the sharp asymptotics. A more detailed explanation of the way that Carlen's 
approach has been modified here is given in the concluding remarks in Section~\ref{sec:conc}.

This paper is organized as follows: In Section~\ref{sec:def} we fix the notation and
give the basic definitions for a quantum spin. 
Then, we recall conjectures concerning Wehrl's entropy and some background 
information in Section~\ref{sec:con}. After that, we present a sharp norm bound
and derive an estimate on the Wehrl entropy, followed 
by proofs in Section~\ref{sec:res}. We conclude with a review of the 
strategies employed here and possible improvements.

\section{Basic definitions}\label{sec:def}

The mathematical description of a spatially fixed, non-relativistic 
spinning quantum particle involves a $(2j+1)$-dimensional Hilbert
space, where $j\in\frac{1}{2}\N$ is called the spin quantum number.
One may identify this space with the complex Euclidean space $\C^{2j+1}$
equipped with the canonical inner product. The symmetry group describing
the purely internal quantum degree of freedom is SU(2), and its representation
on $\C^{2j+1}$ may be constructed by an inductive procedure \cite{Lee88,Sch99}
from that on $\C^2$.
An alternative to this construction is based on choosing a function space
related to a coherent-state resolution of the identity \cite{Kla59,Ber74,Per86}. This 
function space
incorporates a correspondence principle and makes contact with a classical
mechanical system having the sphere as its phase space.
In the co-ordinates we use, the functions are defined on 
the Riemann sphere, \ie the compactified complex plane. Elements of SU(2) 
act in close analogy with rotations, realized as Moebius transformations 
in the complex plane.

\begin{definition}
  Each vector $f$ in the complex linear space $\F_j$ is given as the product of a 
conformal factor and a polynomial of degree at most $2j$,
\begin{align} 
    f: \: \C \rightarrow \C, \:\:
        z \mapsto \frac{1}{(1+\abs{z}^2)^j}\sum_{k=0}^{2j} c_k z^k
\end{align}
with complex coefficients $c_k\in \C$, $k\in\{0,1,\dots,2j\}$.
We define an inner product $\langle f,g\rangle$ between vectors $f$ and $g$
as
\begin{equation}
  \langle f,g\rangle := 
   \frac{2j+1}{\pi}\int_{\C} \ol{f(z)} g(z) \frac{d^2z}{(1+\abs{z}^2)^2} \, ,
\end{equation}
by convention conjugate linear in the first entry.
\end{definition}

\begin{remarks}
The elements of the group SU(2) may be thought of as matrices 
$(\alpha ,-\ol \beta; \beta ,\ol \alpha)$ that are specified 
by two complex
parameters $\alpha,\beta \in \C$ satisfying $\abs{\alpha}^2+\abs{\beta}^2=1$.
The group acts on a function $f$  in the space $\F_j$ by Moebius transformations of
the argument combined with a unimodular multiplier,
\begin{equation}
   T_{\alpha,\beta} f(z) 
   = \frac{(\beta z + \ol \alpha)^{2j}}{\abs{\beta z + \ol \alpha}^{2j}}
     f\Bigl(\frac{\alpha z-\ol \beta}{\beta z + \ol \alpha}\Bigr) \, .
\end{equation}
One may verify that indeed the inner product of $\F_j$ is invariant under such transformations 
$\{T_{\alpha,\beta}\}$,
because $\frac{d^2z}{(1+\abs{z}^2)^2}$ is the rotation-invariant measure on the sphere expressed
in stereographic coordinates.

The space $\F_j$ is equipped with a reproducing kernel $K$,
\begin{equation}
  K(z,w) = \frac{(1+z\ol w)^{2j}}{(1+\abs{z}^2)^j(1+\abs{w}^2)^j} 
\end{equation} 
that yields
\begin{equation}
  \label{eq:pev}
  \langle K(\cdot,w),f\rangle=f(w) 
\end{equation}
for all $f \in \F_j, w \in \C$.
The functions $\{K(\cdot,w)\}\subset \F_j$, indexed by $w \in \C\cup\{\infty\}$ with the
convention $K(z,\infty)= \frac{z^{2j}}{(1+\abs{z}^2)^j}$, are also known
as coherent vectors. 
\end{remarks}

\section{Conjectured norm and entropy bounds}\label{sec:con}

A correspondence principle \cite{Ber74} suggests viewing the compactified complex 
plane as the phase-space of a classical system. From this point of view, the
definition of the Hilbert space $\F_j$ implies an uncertainty principle.
Unlike in $L^2$-spaces, one cannot arbitrarily concentrate the contribution 
to the norm of a function $f\in\F_j$ around a given point.

\begin{definition}
The vectors in $\F_j$ are by their definition all bounded functions and thus
contained in the $L^p(\SS)$ spaces on the Riemann sphere $\SS=\C\cup\{\infty\}$. 
For each $p \ge 1$, we define the 
normalized $p$-norm of $f$ as
\begin{equation}
  \label{eq:nnorm}
  \nnorm{f}_p := \left(\frac{pj+1}{\pi} \int_\C \abs{f(z)}^p \dz \right)^{1/p} \, .
\end{equation}
\end{definition}

\begin{remarks}
  Since the underlying measure is invariant under Moebius transformations, these $p$-norms 
are all invariant under the action of SU(2), 
\begin{equation}
   \nnorm{T_{\alpha,\beta}f}_p=\nnorm{f}_p
\end{equation}
where again $\alpha,\beta\in\C$ satisfy $\abs{\alpha}^2+\abs{\beta}^2=1$.

The unusual, explicit $p$-dependence via the prefactor $pj+1$ in 
the definition \eq{nnorm} is chosen in order
to have $\nnorm{K(\cdot,w)}_p=1$ independent of $w \in \C\cup\{\infty\}$ and $p\ge 1$.
This may be verified by explicit calculation after using an appropriate $T_{\alpha,\beta}$ 
rotating the index $w$ to the origin.
By the point-evaluation property \eq{pev}, the Cauchy-Schwarz inequality and 
the normalization of coherent vectors, we estimate
$\abs{f(z)}\leq\nnorm{f}_2$. Therefore,
when $f$ is normalized according to $\nnorm{f}_2=1$,
the probability density $\rho_f: z \mapsto \abs{f(z)}^2\leq 1$ cannot be arbitrarily concentrated, 
which provides a first albeit crude uncertainty principle.
\end{remarks}

\begin{definition}
The Wehrl entropy of $f\in\F_j$, normalized according to $\nnorm{f}_2=1$, 
is defined as
\begin{equation}
  S_j(|f|^2) := - \frac{2j+1}{\pi} \int_\C \abs{f(z)}^2 \ln \abs{f(z)}^2 \dz \, .
\end{equation}
\end{definition}

\begin{remark}
By the pointwise estimate in the preceding remark, $S_j$ is seen to be non-negative. This may also
be proved using Jensen's inequality \cite{Lie78}. Coherent vectors are believed to be the most 
concentrated wavepackets and therefore expected to minimize $S_j$ and various other measures of 
uncertainty. This is suggestive in the light of the point evaluation property \eq{pev}, which is 
in sufficiently large function spaces solely achieved by integrating against Dirac's delta function. 
\end{remark}

The somewhat vague notion of coherent vectors being the ``most concentrated'' motivates the following 
conjecture in the spirit of Lieb's approach to the Wehrl entropy bound \cite{Lie78}.

\begin{conjecture}
The normalized $p$-norm of a given $f\in \F_j$ is decreasing in $p\ge 1$.
More precisely, 
\begin{equation}
    \label{eq:nest}
      \nnorm{f}_q \leq \nnorm{f}_p \, 
\end{equation}
for all $q \ge p \ge 1$, and when $q>p$, equality holds if and only 
if $f$ is collinear with a coherent vector
\begin{equation}
  f(z) = c K(z,w)
\end{equation}
for some fixed pair $c \in \C$ and $w \in \C\cup \{\infty\}$.
\end{conjecture}

A related conjecture has recently been discussed in the literature 
in the context of measures of localization in phase-space, see \cite{GZ01}
and references therein. It appears there as a conjectured 
lower bound for the so-called R\'enyi-Wehrl entropy. In terms of our 
notation, it amounts to stating the following special case of the 
above conjecture.


\begin{conjecture}
Specializing to $p=2$ in the preceding conjecture and using
the monotonicity properties of the $q$-th
power and of the logarithm, \eq{nest} implies
for $f\in \F_j$ with $\nnorm{f}_2=1$ and any $q>2$ that
\begin{equation}
   \label{eq:npos}
   \frac{2}{2-q} \ln(\nnorm{f}_q^q)\ge 0 \, ,
\end{equation}
with equality again if and only if $f$ is collinear with a coherent vector.
\end{conjecture} 

We henceforth refer to the left-hand side of the above inequality
as an entropy of R\'enyi-Wehrl type of index $q/2$, but caution 
the reader that it differs from the definition of the usual R\'enyi-Wehrl 
entropy in the literature \cite{GZ01,Sug02} by the explicit $q$-dependence 
contained in the norm $\nnorm{f}_q$. A virtue of the normalization 
used here is that neither \eq{nest} nor \eq{npos} are explicitly $j$-dependent.
In addition, the conjectured monotonicity in \eq{nest} and non-negativity in \eq{npos} 
are notationally identical with properties of R\'enyi's entropy in discrete measure spaces, 
see \cite{Zyc03} or equivalent statements in \cite[Section 5.3]{BS93} in terms of the R\'enyi information.
The reason for this similarity is that the minimizers of R\'enyi's entropy in the 
discrete measure spaces are normalized with respect to all $l^q$-norms, just as the coherent
vectors are in our case. One may wonder whether the counterparts of other properties described 
in \cite{Zyc03} or \cite{BS93}, such as convexity and monotonicity of 
$\frac{2}{2-q}\ln(\nnorm{f}_{q}^{q})$ in $q$, are also satisfied.

The analogue of \eq{npos} has been shown to hold when $q$ is an even 
integer in a rather general setting \cite{Sug02}, thereby extending 
a tensorization argument for SU(2), see \cite{Lee88} or \cite[Theorem 4.3]{Sch99}, 
to compact semisimple Lie groups, see also the related results \cite{Sai80,Bur83,Bur87,Luo97}. 
However, so far we still lack a proof that includes those $q$ that are arbitrarily 
close to $p=2$. If those were included in a proof of either of the preceding conjectures,
then the following bound on the Wehrl entropy would result by endpoint differentiation. 

\begin{conjecture}[Lieb, 1978]
For all $f \in \F_j$ with $\nnorm{f}_2=1$, the Wehrl entropy is bounded below by
\begin{equation}
  \label{eq:liebcon}
  S_j(|f|^2) \ge \frac{2j}{2j+1} 
\end{equation}
and equality holds whenever $f$ is up to a unimodular constant given by 
a coherent vector.
\end{conjecture} 

\section{Results}\label{sec:res}

The first result presented in this section is that for a given $f\in\F_j$, the norm 
$\nnorm{f}_q$ is decreasing when restricted to a discrete set of values for $q$.
A special case of this result implies via interpolation and endpoint differentiation
the bound on the Wehrl entropy with sharp asymptotics.

\begin{theorem}\label{thm:1}
The conjectured norm estimate \eq{nest} holds
for all $f \in \F_j$ if $q=p+\frac n j$ with any integer $n \in \N$ and 
$p > \frac 1 j$. The estimate is sharp because equality holds in \eq{nest} if $f$ is 
collinear with a coherent vector.
\end{theorem}
\begin{consequence}
Specializing to $p=2$ and $\nnorm{f}_2=1$,
this estimate implies via the monotonicity properties of the logarithm and 
of the $q$-th power that \eq{npos} is true for $f\in F_j$ when 
$q=2+\frac n j$ with $n\in \N$.
\end{consequence}

Unlike in previous results \cite{Sai80,Bur83,Bur87,Luo97,Sch99,Sug02}, 
the index $q/2$ is no longer restricted to integer values.
Another, less direct consequence of Theorem~\ref{thm:1} 
is the following bound on Wehrl's entropy. 

\begin{theorem}\label{thm:2}
The Wehrl entropy associated with $f \in \F_j, \nnorm{f}_2=1$, is bounded below
by 
\begin{equation}
  \label{eq:thm2}
   S_j(|f|^2)\ge 2j \ln\Bigl(1+\frac{1}{2j+1}\Bigr) \, .
\end{equation}
\end{theorem}

\begin{remarks}
This bound has the conjectured high-spin asymptotics up to, but not including, first 
and higher order terms in $j^{-1}$, because
\begin{equation}
  0 \leq \frac{2j}{2j-1}-2j\ln\Bigl(1+\frac{1}{2j+1}\Bigr)
    = \frac{2j}{(2j+1)^2}\int_0^1\frac{x}{1+\frac{x}{2j+1}}\,dx < \frac{1}{4j} \, .
\end{equation}
Since highest-weight SU(2) coherent states
approach Glauber coherent states when scaled appropriately in the limit 
$j\to\infty$ \cite{ACGT72}, see also \cite{BLW99}, 
one could use these asymptotics to reproduce the entropy bound related to Glauber coherent 
states obtained by Lieb \cite{Lie78}. However, the route via finite dimensional spaces 
and SU(2) does not seem to offer the vast simplification that was originally expected.

By the usual concavity argument \cite{Weh79,Lie78,Sch99}, the Wehrl entropy bound derived here
may be extended to include all quantum states, not only pure ones.
To this end, we only need to replace the density $\rho_f=|f|^2$ that appears in the definition of 
$S_j$ by the more general $\rho: z \mapsto \sum_k \abs{f_k(z)}^2$ with any orthogonal family
$\{f_n\}_{k=1}^{2j}\subset \F_j$ that satisfies $\sum_k \nnorm{f_k}_2^2=1$.
Another possible generalization of the bound \eq{thm2} is to include several degrees of 
freedom using the monotonicity of Wehrl's entropy \cite{Weh79,Lie78}. 
\end{remarks}

\subsection{Proof of Theorem~\protect\ref{thm:1}}

\begin{proof}
To begin with, we remark that it is enough to show the result for $n=1$ and $q> p>q/2$, 
otherwise we simply iterate the inequality.
The strategy for the proof of Theorem~\ref{thm:1} is as follows: First, we restate the 
norm bound \eq{nest} as the result of an optimization problem involving the quadratic form of 
the Laplacian on the sphere. This step relies on identity~\eq{car} that is derived in 
the spirit of Carlen \cite{Car91} from the special properties of the
function space $\F_j$. 
Then, we enlarge the space when looking for optimizers without losing the
sharpness of the bound, because coherent vectors are optimal functions. 

For the detailed explanation, it is convenient to introduce 
the norm $\norm{f}_p=\nnorm{f}_p/(pj+1)^{1/p}$ of $f\in L^p(\SS)$ with respect to 
the rotationally-invariant probability measure on $\SS$. Using the Carlen Identity 
\eq{car}, we have for $q\ge p > 0$
\begin{equation}
  \label{eq:step1}
  \max_{f\in\F_j\setminus\{0\}} \frac{\norm{f}_q^q}{\norm{f}_p^q}
  = \max_{f\in\F_j\setminus\{0\}} \frac{\norm{f}_q^q
                           -\frac{4}{qj(qj+2)}\int 
                                                 \abs{\partial 
                                                       \abs{f}^{q/2}
                                                     }^2 \frac{d^2z}{\pi}                          
                           }{(1-\frac{1}{qj+2})\norm{f}_p^q} \, .
\end{equation}
Without loss of generality, we can assume that each
$f$ vanishes at infinity, $\lim_{z \to \infty}$ $f(z)=0$. Either this assumption 
is satisfies right away, or the polynomial part of $f$ has highest 
degree and thus it has a zero in the complex plane that can be rotated to infinity by a 
suitable Moebius transformation without changing any of the norms in \eq{step1}.
Now we change notation to $u=\abs{f}^{q/2}$ and enlarge the set of such $f$, 
\begin{equation}
 \max_{f} \frac{\norm{f}_q^q}{\norm{f}_p^q}
 \leq \sup_u \frac{\norm{u}_2^2
                           -\frac{4}{qj(qj+2)}\int 
                                                 \abs{\partial u}^2 \frac{d^2z}{\pi}     
                           }{(1-\frac{1}{qj+2})\norm{u}^2_{2p/q}} \label{eq:maxim}
\end{equation}
where the supremum is taken over the set $\{0\leq u\not\equiv 0\}$ 
in the space $C_0^1(\C)$ of bounded, continuously differentiable functions on the plane
that vanish at infinity.
In Lemma~\ref{lem:exist} we see that the supremum in \eq{maxim} has a finite value
and even deduce the existence of a maximizing function for this supremum.
Given any maximizer, performing a spherically symmetric decreasing rearrangement
must necessarily preserve its gradient norm. Therefore, we may restrict
ourselves to proving in
Lemma~\ref{lem:uniq} that in the class of rotationally symmetric functions,
this maximizer is unique, up to an overall positive constant. It satisfies the 
Euler-Lagrange variational equation
\begin{equation}
  \label{eq:ELeqn}
  \Bigl(1+\frac{4}{qj(qj+2)}\Delta\Bigr) u = b u^{\frac{2p}{q}-1}
\end{equation}
with $\Delta = \frac 1 4 (1+\abs{z}^2)^2(\partial_1^2+\partial_2^2)$ 
being the spherical Laplacian and a constant $b>0$ that is fixed 
by choosing the norm $\norm{u}_{2p/q}$.
By inspection, this unique function is seen to be $u: z \mapsto A (1+\abs{z}^2)^{qj/2}$ 
with $A=((qj+2)b/qj)^{q/2(q-p)}$,
and thus the corresponding $f$ is up to a constant factor a coherent vector.

Consequently, inequality \eq{nest}  
follows from reverting to the original normalization convention. It is sharp
since equality is achieved if $f$ is a coherent vector.
\end{proof}

\subsection{Proof of Theorem~\protect\ref{thm:2}}

This is merely a consequence of Theorem~\ref{thm:1}. It follows from interpolating the
norm bound \eq{nest} and endpoint differentiation.

\begin{proof}
Let us choose $q=2+\frac 1 j$ in Theorem~\ref{thm:1} and for
$2<s<q$ select $\theta>0$ such that $\frac 1 s = \frac{1-\theta}{2}+\frac \theta q$. 
Using the same notation $\norm{f}_s=\nnorm{f}_s/(sj+1)^{1/s}$ as before, 
H\"older's inequality combined with the norm bound \eq{nest} gives 
\begin{align}
  \norm{f}_s &\leq \norm{f}_2^{1-\theta} \norm{f}_q^\theta \\
             &\leq \Bigl[ \frac{(2j+1)^{1/2}}{(qj+1)^{1/q}} \Bigr]^\theta \norm{f}_2 \, ,
\end{align}
so eliminating $\theta$ results in
\begin{equation}
  \norm{f}_s^s \leq \frac{(2j+1)^{\frac q 2 \frac{2-s}{2-q}}}{(qj+1)^{\frac{2-s}{2-q}}} \norm{f}_2^s \, .
\end{equation}
At $s=2$ both sides are equal, thus one may differentiate
\begin{align}
   & \left. s \frac{d}{ds}\right\vert_{s=2} \norm{f}_s^s =
   \frac 1 \pi \int_\C \abs{f(z)}^2 \ln \abs{f(z)}^2 \dz \label{eq:logderiv}\\
  & \leq s \left. \frac{d}{ds}\right\vert_{s=2} 
                  \frac{(2j+1)^{\frac q 2 \frac{2-s}{2-q}}}{(qj+1)^{\frac{2-s}{2-q}}}
         = \ln(2j+1) - 2j \ln\Bigl(\frac{2j+2}{2j+1}\Bigr) 
\end{align}
and by a change in normalization and an overall change of sign arrives at the desired entropy bound~\eq{thm2}.
The justification for differentiating under the integral sign in \eq{logderiv} is dominated convergence and 
the estimate $0\ge 2 (\abs{f}^{s-2}-1)/(s-2)\ge\ln\abs{f}^2$. 
\end{proof}

\subsection{Ingredients of the proof of Theorem~\protect\ref{thm:1}}

\begin{lemma}[Carlen Identity]
For $q>0$ and all $f \in \F_j$, $j \in \frac 1 2 \N$, the following identity is true
\begin{equation}\label{eq:car}
  \int_\C \abs{\partial \abs{f}^{q/2}}^2 \frac{d^2z}{\pi}
 = \frac{q j}{4\pi} \int_\C \abs{f}^{q} \dz \, .
\end{equation}
\end{lemma}
\begin{proof}
We will first prove a regularized version of the identity. Let $\epsilon>0$ and
assume $\phi: \C \to \C$
is a complex polynomial of maximal degree $2j$, we will show for $q>0$ that
\begin{align}
  \int_\C \abs{\partial 
               \frac{(\abs{\phi(z)}^2+\epsilon)^{q/4}}{(1+\abs{z}^2)^{q j/2}}
              }^2 \frac{d^2z}{\pi} 
 = \frac{q j}{4} \int_\C 
              \frac{ (\abs{\phi(z)}^2+\epsilon)^{q/2}}{(1+\abs{z}^2)^{q j +2}} 
         \frac{d^2z}{\pi} 
    + E(\epsilon)
\end{align}
with an error term
\begin{equation}
  E(\epsilon) = \frac{\epsilon q}{8} \int_\C (\abs{\phi(z)}^2+\epsilon)^{\frac q 2 - 2} 
                \frac{\abs{\partial \phi(z)}^2}{(1+\abs{z}^2)^{q j}} \frac{d^2 z}{\pi}
\end{equation}

This regularized identity is obtained by elementary calculus operations involving the 
complex derivative $\partial$, observing that $\phi$ is holomorphic, and integration by parts:
\begin{align}
  &\int_\C \abs{\partial 
               \frac{(\abs{\phi(z)}^2+\epsilon)^{q/4}}{(1+\abs{z}^2)^{qj/2}}
              }^2 \frac{d^2z}{\pi}
 = \int_\C \left[
        \frac{q^2}{16} \frac{\abs{\phi(z)}^2 \abs{\partial\phi(z)}^2 
                       (\abs{\phi(z)}^2+\epsilon)^{\frac q 2 - 2}}{(1+\abs{z}^2)^{q j}}
         \right. \\
  &\phantom{=\int} - \frac{q^2 j}{4}  
        \frac{\Re [\ol{\phi(z)} z \partial \phi(z)] 
              (\abs{\phi(z)}^2+\epsilon)^{\frac q 2 -1}}{(1+\abs{z}^2)^{q j +1}} 
   \left.+ \frac{q^2 j^2}{4} 
      \frac{\abs{z}^2 (\abs{\phi(z)}^2+\epsilon)^{qj/2}}{(1+\abs{z}^2)^{qj +2}} 
        \right] \frac{d^2z}{\pi} \nonumber\\
  &= \int_\C \left[ \frac{1}{\conf{qj}} \bigl( \frac 1 4 \partial \ol \partial \reg{q/2} +
                                                 \epsilon \frac q 8 \reg{\frac q 2 -2} 
                                                 \abs{\partial \phi(z)}^2 \bigr)\right.\\
  &  \phantom{=\int}- \frac{qj}{2} \frac{\Re [z \partial \reg{q/2}]}{\conf{qj +1}} 
     \left. + \frac{q^2 j^2}{4} \frac{ \abs{z}^2 \reg{q/2}}{\conf{qj +2}} \right] \frac{d^2z}{\pi}
           \nonumber\\
  & = \int_\C \reg{q/2} \left[ \frac 1 4 \partial \ol \partial \frac{1}{\conf{q j}}
                              +\epsilon\frac q 8 \frac{\abs{\partial\phi(z)}^2}{\reg{2}} \right.\\
    & \phantom{=\int}\left.- \frac{qj}{2} \Re\bigl[\partial \frac{z}{\conf{q j +1}}\bigr] 
      + \frac{q^2 j^2}{4} \frac{\abs{z}^2}{\conf{q j+2}}\right] \frac{d^2z}{\pi} \nonumber\\
 & = \frac{q j}{4}  \int_\C \frac{\reg{q/2}}{\conf{qj+2}}\frac{d^2z}{\pi}
    +\epsilon \frac{q}{8} \int_\C \frac{\reg{\frac q 2 - 2}\abs{\partial\phi(z)}^2}{\conf{q j}}
    \frac{d^2z}{\pi}  \, .
\end{align}
There are no boundary terms in the integration by parts because the growth of the powers of
$\abs{\phi}$ and of their derivatives are suppressed by sufficiently strong polynomial growth 
in the denominator.

In the limit $\epsilon\to 0$, the error term vanishes by dominated convergence,
yielding identity~(\ref{eq:car}).
Dominated convergence applies because either $q\ge 4$ and  
$(\abs{\phi(z)}^2+\epsilon)^{\frac q 2 -2}\leq 
C(1 + \abs{z}^2)^{j(q-4)}$ with a fixed constant $C>0$ valid for all sufficiently small $\epsilon$, 
or $0<q<4$ and we estimate on the set $\{z: \abs{\phi(z)}^2<\epsilon\}$ the expression
$\epsilon (\abs{\phi(z)}^2+\epsilon)^{\frac q 2 -2}
  \leq 2 \frac{(4-q)^{2-\frac q 2}}{(2-q)^{3-\frac q 2}} \abs{\phi(z)}^{q-2}$. 
To ensure the validity of dominated convergence, one may then recall that $\phi$ has
only isolated zeros of finite order.  
\end{proof}

\begin{remark}
A version of the identity \eq{car} was first shown by Carlen \cite{Car91} for
the function space related to Glauber coherent states. One may recover his result
by the usual group contraction procedure in the limit $j\to \infty$ while
scaling $z\to z/\sqrt{2j\hbar}$ with $\hbar>0$.

The case $q=2$ can
be verified in a simpler way than the explicit calculation given here. All
that is needed is to verify that the square of the gradient norm in \eq{car}
defines a quadratic form on $\F_j$. Appealing to Schur's lemma, the corresponding
operator is a multiple of the identity, because the quadratic form is 
invariant under the irreducible unitary representation of SU(2). To obtain the
correct numerical factor, one may then simply choose $f$ to
be the coherent vector centered at the origin and evaluate the left-hand side
of equation~\eq{car}.
\end{remark}

\begin{lemma} \label{lem:exist}
The supremum of the expression \eq{maxim} is attained for some function $u$.
In other  words, there is some $u$ with $\norm{u}_s=1$, $1<s<2$,
satisfying $0\leq u$ and $u(z)\to 0$ as $z \to \infty$, which maximizes
\begin{equation}
  \label{eq:maxim2}
   \norm{u}_2^2-\frac{4}{qj(qj+2)}\int_\C \abs{\partial u}^2 \frac{d^2z}{\pi} \, .
\end{equation}
\end{lemma}
\begin{proof}
We abbreviate $s:=\frac{2p}{q}$ and denote the dual index as $s':=\frac{s}{s-1}$.
The Sobolev inequality on the two-sphere states 
\begin{equation}
  \norm{u}_2^2+ \frac{s'-2}{2} \norm{\nabla u}_2^2 \ge \norm{u}_{s'}^2 \, .
\end{equation}
Note that for real-valued $u$, in  two dimensions the two expressions
$\norm{\nabla u}^2_2$ and $\int_\C \abs{\partial u}^2 d^2z/\pi$
are identical, one using the gradient $\nabla$ on the sphere and the other
the complex derivative $\partial=(\partial_1-i\partial_2)/2$. Therefore,
employing the Sobolev and H\"older inequalities yields for $\norm{u}_{s}=1$
a finite bound
\begin{align}
  \label{eq:bnd}
  &\norm{u}_2^2-\frac{4}{qj(qj+2)}\int_\C \abs{\partial u}^2 \frac{d^2z}{\pi} \nonumber \\
  &\leq \Bigl( 1 + \frac{2}{r-2}\frac{4}{qj(qj+2)}\Bigr) \norm{u}_s\norm{u}_{s'}
                                  -\frac{2}{r-2}\frac{4}{qj(qj+2)} \norm{u}^2_{s'}\\
  &\leq \max_{x\ge 0} \Bigl[\Bigl( 1 + \frac{2}{r-2}\frac{4}{qj(qj+2)}\Bigr) x - 
                           \frac{2}{r-2}\frac{4}{qj(qj+2)} x^2 \Bigr] \, . \label{eq:yeah}
\end{align}

Now assume a maximizing sequence $(u_n)_{n\in\N}$ such that $0\leq u_n$ and 
$\norm{u_n}_s=1$ for all $n\in\N$.
Apart from the finiteness of the supremum over $s$-normalized $u$ in \eq{bnd}, 
inequality \eq{yeah} shows that the sequence $(x_n)_{n\in\N}$ given by 
$x_n:=\norm{u_n}_{s'}$ is bounded, and because of the compactness
of the sphere, the same bound applies to the $2$-norms of the 
sequence $(u_n)_{n\in\N}$. This, in turn, 
forces any maximizing sequence to be uniformly bounded in gradient $2$-norm.

We may now without loss of generality replace each $u_n$ by its equimeasurable 
symmetric decreasing rearrangement, since this only decreases the gradient norm by the
contractivity properties of the heat semigroup on the sphere, just as
in the Euclidean case \cite[Lemma 4.1]{Lie83}.
In addition, we can use Helly's selection principle, that is, choose a subsequence 
that converges on all rational radii. By the monotonicity of each $u_n$, the convergence 
extends to almost every radius. Appealing to a Rellich-Kondrakov compact embedding
theorem \cite[Theorem 2.34]{Aub82}, the limit $u_n\to u$ is in the $s$-norm. 
The compact embedding we use is that of the Sobolev space $H_1^1(\SS)$ in $L^s(\SS)$.
\end{proof}

\begin{lemma} \label{lem:smooth}
Given $s>1$, every maximizer $u\in L^s(\SS)$ 
in the preceding lemma is a smooth function.
\end{lemma}
\begin{proof}
  From calculus of variations, any maximizer $u\in L^s(\SS)$ is a distributional solution of 
the corresponding Euler-Lagrange equation that has the form of a Poisson equation,
\begin{equation}
  \Delta u = a u^{s-1} - b u
\end{equation}
with constants $a,b>0$.
By the integrability of the Green function $G$ on the sphere,
in complex coordinates given as
\begin{equation}
  G(z,w) = - \ln\left(\frac{4 \abs{z-w}^2}{(1+\abs{z}^2)(1+\abs{w}^2)}\right) \, ,
\end{equation}
and the identity
\begin{equation}
  u(w) = \frac 1 \pi \int_{\C} G(w,z)(au^{s-1}(z)-bu(z)) \dz
\end{equation}
we see that $u$ is bounded and continuous. 

In fact, its first derivative is seen to be H\"older continuous with any index $0<\alpha<1$
from replacing $u$ with a directional derivative and calculating the magnitude of the 
gradient of the Green function 
\begin{equation}
  \abs{\nabla G(z,w)} = \left( \frac{(1+\abs{z}^2)(1+\abs{w}^2)}{\abs{z-w}^2}
                               - \frac{1}{(1+\abs{z}^2)(1+\abs{w}^2)}\right)^{1/2} \, .
\end{equation} 

By bootstrapping, $u$ is smooth since its $k$-th derivative is
H\"older continuous with index $0<\alpha_k<(s-1)^{k-1}$.
\end{proof}

\begin{lemma} \label{lem:uniq}
Given $q>p>q/2$, there is a unique rotationally symmetric 
non-trivial solution $u$ to the Euler-Lagrange equation 
\eq{ELeqn} that meets the non-negativity and limit requirements $u\ge0$ and 
$\lim_{z\to\infty}u(z)=0$.
\end{lemma}
\begin{proof}
We know from the preceding lemma that $u$ is differentiable, even smooth.
Therefore, a rotationally symmetric $u$ considered as a function 
of $r=\abs{z}$ solves the ordinary differential equation 
\begin{equation}
  \label{eq:ODE}
  (1+r^2)^2 \frac 1 r (r u'(r))' -a (u(r))^{\frac{2p}{q}-1}+b u(r) = 0
\end{equation}
with an initial condition $u'(0)=0$ and an unknown value 
$u(0)>0$. We will abbreviate the non-linearity as 
$\phi(u):=-bu^{(2p-q)/q}+u$. Despite the singularity at $r=0$,
the initial values $u(0)$ and $u'(0)=0$ uniquely determine a solution
on all $\{r\ge 0\}$. This follows with the help of the
Schauder fixed point theorem as in \cite[Appendix]{FLS96}
by replacing the expression for $u'(r)$ in the Euclidean
case treated there with \eq{int1} given hereafter. To show the claimed uniqueness
of a solution meeting the non-negativity and limiting requirements,
we first show that any such solution is
\textit{strictly} radially decreasing. Then, we
exclude the cases of multiple solutions having infinitely many,
finitely many, and finally having no intersections. In the
literature, such arguments have been called separation lemmas.
We adapt elements of \cite{PS83,PS86,FLS96,ST00} to the setting of the
sphere.

To begin with, we show that $u$ is strictly decreasing as a function
of $r$. Integrating the ODE \eq{ODE} gives
\begin{equation}
  \label{eq:int1}
  u'(r) = - \frac{1}{r} \int_0^r \frac{t}{(1+t^2)^2} \phi(u(t))\, dt \, .
\end{equation}
If there some critical point $r_0>0$ for $u$, then we could conclude that
both sides of this equation vanish and thus, inserting \eq{int1} simplifies \eq{ODE} to
\begin{equation}
  u''(r_0)=-\frac{1}{(1+r_0^2)^2} \phi(u(r_0))=0 \, .
\end{equation}
From the uniqueness of solutions to initial value problems having 
Lipschitz-continuous
coefficients we conclude that $u$ is necessarily the non-zero constant
determined by $\phi(u)=0$, contradicting the limiting requirement 
in our assumption. Therefore, $u$ cannot have any critical points 
for $r>0$.

Now we exclude the three possible cases of multiple solutions.
\begin{enumerate}
\item
{\it Two solutions with infinitely many intersections.}
Assuming there are two solutions $u$ and $v$ that intersect infinitely often, then
there are radii $0<r_0<r_1$ such that
\begin{enumerate}
\item $u(r_1) = v(r_1)$, $u'(r_0) = v'(r_0)$, and $0>v'>u'$ on $(r_0,r_1)$,
\item $\phi(u(r))<0$ and $\phi(v(r))<0$ for all $r\ge r_0$.
\end{enumerate}
Define $\Phi(u):=\int_0^u \phi(x)\, dx$. By comparison with energy dissipation
in a mechanical system, we have the ``conservation law'', here in terms of
the solution $u$,
\begin{align}
 &\left[ \frac 1 2 (1+r^2)^2(u'(r))^2 + \Phi(u(r)) \right]_{r_0}^{r_1}\nonumber\\
 &\phantom{lotsa space} 
   + \int_{r_0}^{r_1}\left(\frac{(1+r^2)^2}{r}-r(1+r^2)\right) (u'(r))^2 \, dr = 0 \, .
\end{align}
Subtracting this identity for the two solutions $u$ and $v$, we obtain
\begin{align}
  &\frac 1 2 (1+r_1^2)^2((u'(r_1))^2-(v'(r_1))^2) + \Phi(v(r_0))-\Phi(u(r_0))\nonumber\\
  & + \int_{r_0}^{r_1}\left(\frac{(1+r^2)^2}{r}-r(1+r^2)\right)((u'(r))^2-(v'(r))^2)\, dr = 0
\end{align}
but the left-hand side is strictly positive in each difference term, 
so this yields a contradiction.
\item \label{it:case2}
{\it Two solutions having finitely many intersections.}
For this part, we choose the parametrization in terms of the azimuthal angle $\theta$,
$r=\tan \theta/2$. In this variable, the ordinary differential equation \eq{ODE}
is expressed as
\begin{equation}
  \label{eq:ODEtheta}
  u''(\theta)+\cot\theta\, u'(\theta) + \phi(u(\theta))=0 \, .
\end{equation}
From the two solutions $u$ and $v$ let us pass to the inverse functions
denoted as $\theta=u^{-1}$ and $\sigma=v^{-1}$.
If there are finitely many intersections, then there is $u_0>0$ such that
$\theta>\sigma>\pi/2$ and $0>\theta'>\sigma'$ on $(0,u_0)$.

Define the difference of ``kinetic energy'' as
\begin{equation}
  B(u) = \frac{1}{(\theta'(u))^2}-\frac{1}{(\sigma'(u))^2}>0 \, ,
\end{equation}
then $B(0)=0$. However, according to the ODE \eq{ODEtheta}, the 
derivative 
\begin{equation}
  B'(u)=-2\left( \frac{\cot \theta}{\theta'}-
                         \frac{\cot\sigma}{\sigma'} \right)<0
\end{equation}
since $0>\cot\theta>\cot\sigma$ and $0>\theta'>\sigma'$. We have
reached a contradiction with the strict positivity of $B$.
\item
{\it Two solutions without intersection.}
As a first step to exclude this case, 
we will show that the difference 
$\theta-\sigma$ of the inverse functions of $u$ and $v$ 
can have at most one critical point,
a local maximum, in an interval where $\theta>\sigma$. 

To this end, we note that in terms of the inverse function $\theta$
of $u$, 
the ODE \eq{ODEtheta} becomes
\begin{equation}
  -\theta''+\cot\theta (\theta')^2 + \phi(u(\theta))(\theta')^3=0 \, .
\end{equation}
Subtracting this identity for the two solutions at a point 
where $\theta'=\sigma'$ and $\theta>\sigma$ gives 
\begin{equation}
  \theta''-\sigma''=(\cot \theta - \cot \sigma) (\theta')^2<0
\end{equation}
by the monotonicity of the cotangent function.

Consequently, assuming $u>v$, then
the difference $\theta-\sigma$ cannot have
any critical point in $(0,v(0))$, because
this would contradict $\theta'-\sigma'\to \infty$ as $u\to v(0)$.
So $\theta'-\sigma'>0$ in $(0,v(0))$. 
But similarly as in the preceding case of finitely many intersections, 
this is impossible near $u=0$, because in terms of the azimuthal angle 
we would then have $B(0)=0$, $B'(u)<0$ for small $u$. 
\end{enumerate}
\end{proof}

\section{Conclusion}
\label{sec:conc}

The lack of a sharp entropy bound remains frustrating. 
Many promising attempts have failed. Among others, this includes
the hope that group-theoretic, explicit computations as in \cite{Sch99} may give 
rise to an inductive argument; the use of the Hardy-Littlewood-Sobolev
inequality on the sphere that may be found in so many uncertainty principles 
\cite{Bec93,Bec95,Bec01};
or  via a change of weight analogous to \cite{Luo00} the attempt to
combine variants of the Carlen identity with a logarithmic 
Sobolev-type inequality on the sphere that 
breaks down because of a domain problem \cite[Section 5]{Gro99}. 
At this point the best bet may still be the difficult task of finding sharp 
smoothing properties of the Berezin transform. For lack of a manageable alternative, 
we replaced the Berezin transform with
the differential operator $1+\frac{4}{qj(qj+2)}\Delta$, which should intuitively 
be a good approximation for high spin quantum numbers. 
This particular choice of approximation is motivated by considering optimization 
problems of the form \eq{maxim} and by demanding that coherent vectors solve the corresponding 
Euler-Lagrange equation. The downside of replacing the Berezin transform
is that we cannot characterize the cases
of equality in Theorem~\ref{thm:1} because there is no strong form 
of the Riesz theorem for the gradient norm under radially symmetric decreasing rearrangements.
We cannot resolve this deficiency by invoking a duality principle as in \cite[Theorem 6]{Bec01},
because the operator $1+\frac{4}{qj(qj+2)}\Delta$ is not invertible.

One may expect that an adaptation of the techniques employed here yields
analogous bounds for the Wehrl entropy in the setting of SU(1,1) 
as well, see \cite{Luo97}. It could be worthwhile to include
other highest-weight representations \cite{GZ01}. However, the most important task
is still to find the road that will finally lead to the sharp bounds. Most likely it 
will not be a simple group-theoretic argument, but all of the partial results 
so far point to the very geometric nature and inherent beauty of the problem. 
Discovering these facets of entropy makes the difficulties in resolving Lieb's 
conjecture much less frustrating and all the more fascinating.\\

\paragraph{\bf Acknowledgments}
Thanks go to Elliott Lieb for remarks that were always to the point
and for suggesting this incredibly beautiful and challenging open problem to me. 
I am also grateful for helpful and motivating conversations with
Shannon Starr, Michael Aizenman, Robert Seiringer, 
Almut Burchard, Stephen B. Sontz and Bob Sims.  The referee is acknowledged
for pointing out analogies between the material presented here and
classical results for the R\'enyi entropy on discrete measure spaces.
This work was partially supported under the National Science Foundation grant
PHY-0139984.


\end{document}